# Retrieval of Chemical Abundances in Titan's Upper Atmosphere from Cassini UVIS Observations with Pointing Motion

**Siteng Fan[1], Donald E. Shemansky[2], Cheng Li[1], Peter Gao[3], Linfeng Wan[4], Yuk L. Yung[1,5]**

[1]California Institute of Technology, Pasadena, CA, 91125.

[2]Space Environment Technologies, Altadena, CA, 91101

[3]University of California, Berkeley, CA, 94720.

[4]University of California, Santa Cruz, CA, 95064.

[5]Jet Propulsion Laboratory, California Institute of Technology, Pasadena, CA, USA

Corresponding author: Siteng Fan (stfan@gps.caltech.edu)

**Key Points:**

- An innovative method is developed to analyze Cassini UVIS stellar occultation observations.

- Abundances of major hydrocarbon and nitrile species are derived from observations during a flyby with large instrument pointing motion.

- The new method allows exploration of Titan's upper atmosphere over seasons, latitudes, and longitudes.

**Abstract**

Cassini/UVIS FUV observations of stellar occultations at Titan are well suited for probing its atmospheric composition and structure. However, due to instrument pointing motion, only five out of tens of observations have been analyzed. We present an innovative retrieval method that corrects for the effect of pointing motion by forward modeling the Cassini/UVIS instrument response function with the pointing motion value obtained from the SPICE C-kernel along the spectral dimension. To illustrate the methodology, an occultation observation made during flyby T52 is analyzed, when the Cassini spacecraft had insufficient attitude control. A high-resolution stellar model and an instrument response simulator that includes the position of the point source on the detector are used for the analysis of the pointing motion. The Markov Chain Monte-Carlo method is used to retrieve the line-of-sight abundance profiles of eleven species ($CH_4$, $C_2H_2$, $C_2H_4$, $C_2H_6$, $C_4H_2$, $C_6H_6$, HCN, $C_2N_2$, $HC_3N$, $C_6N_2$ and haze particles) in the spectral vector fitting process. We obtain tight constraints on all of the species aside from $C_2H_6$, $C_2N_2$ and $C_6N_2$, for which we only retrieved upper limits. This is the first time that the T52 occultation was used to derive abundances of major hydrocarbon and nitrile species in Titan's upper and middle atmosphere, as pointing motion prohibited prior analysis. With this new method, nearly all of the occultations obtained over the entire Cassini mission could yield reliable profiles of atmospheric composition, allowing exploration of Titan's upper atmosphere over seasons, latitudes, and longitudes.

## 1 Introduction

Stellar occultation observations made by the Cassini Ultraviolet Imaging Spectrograph (UVIS, Esposito et al. 2004) are essential for constraining the photochemistry of Titan's upper atmosphere, where hydrocarbon and nitrile species show spectral features in the ultraviolet (Hörst 2017). During the last decade, vertical profiles of these species have been derived from selected Titan flybys (Shemansky et al. 2005, Liang et al. 2007, Koskinen et al. 2011, Kammer et al. 2013), which have increased our understanding of physical and chemical processes in Titan's atmosphere. Far-UV (FUV) observations are especially important in constraining the abundances of hydrocarbons more complex than methane, as well as nitriles. However, FUV observations from only five (TB, T21, T41i, T41e and T53) out of the tens of flybys have been used for retrievals to date (Shemansky et al. 2005, Koskinen et al. 2011, Capalbo et al. 2016) due to instrument pointing motion, which causes movement of the stellar image on the UVIS detector. Pointing motion is introduced by the spacecraft attitude control system, which frequently triggers thrusters during occultation observations (Chiang et al. 1993). Deadband of attitude control during stellar occultations is set as ±0.5 mrad (Pilinski & Lee, 2009), comparable to that of a spectral pixel (1.0 mrad × 0.25 mrad), which leads to a shift in the spectral structure. Consequently, the traditional method of analyzing extinction spectra obtained by dividing the nonzero optical depth target spectrum by the extinction-free target spectrum becomes inaccurate. The spectral distortion is non-linear due to UVIS internal instrument scattering. Attempts have been made to determine pointing motions for other instruments, e.g., Cassini/VIMS (Maltagliati et al., 2015), and similar issues arose for similar UV instruments when observing stellar occultations of Titan's atmosphere (e.g. Voyager 1 UVS, Vervack et al. 2004). Using an instrument simulator for forward modeling is essential for evaluating spectra.

Our proposed method uses a Cassini/UVIS simulator for forward modeling that combines the pointing information obtained from the SPICE C-kernel (NASA NAIF) and line-of-sight (LOS) chemical abundances in Titan's atmosphere to simulate spectra. The Markov Chain Monte-Carlo (MCMC) method (Foreman-Mackey et al. 2013) is used as the parameter search tool. Using our proposed method, we derive for the first time vertical profiles of hydrocarbon and nitrile species from flyby T52, which shows significant pointing motion during the entire duration of the observation.

The remainder of this paper is organized as follows: in Section 2, the pointing motion during flyby T52 and its effects are demonstrated; detailed methodology is presented in Section 3; analysis of synthetic spectrum for method testing is in Section 4; spectral analysis results and vertical profiles of retrieved species are given in Section 5, together with brief discussions; our conclusions and the implications for applying this method are discussed in Section 6.

## 2 Compensating for pointing motion

The instrument used in this work is the Cassini/UVIS FUV spectrograph. It covers the spectral range between 1115 and 1912 Å (Esposito et al. 2004). We select flyby T52 for demonstrating our new retrieval method, which took place on April $3^{rd}$, 2009, with the stellar source α Eri. The occultation observation during this flyby is divided into two segments due to different integration durations. We analyze the second segment (NASA PDS ID: FUV2009_093_23_55) in this work, which has an integration duration of 1.75 s for each image; the ray tangent height of incoming stellar light above Titan's surface during this segment was within the critical range (0-1500 km) for atmospheric characterization. The geometry of this segment is given in Table 1, computed from the SPICE C-kernel (NASA NAIF).

**Table 1**. Stellar occultation during T52.

|  | Start | Near 1000km | Near 500km | End |
|---|---|---|---|---|
| Time | 2009-093 23:55:27 | 2009-094 00:10:51 | 2009-094 00:16:51 | 2009-094 00:27:49 |
| Latitude (°) | 32.6 | 35.6 | 37.1 | 39.1 |
| Longitude (°) | 328.9 | 319.5 | 312.8 | 292.7 |
| Ray Height (km) | 2347 | 999 | 502 | -290 |

*Latitude and longitude are computed for the impact parameter (radial vector from body center). It varied by a few degrees during the observation due to spacecraft motion.

The image of α Eri is a point source compared to the pixel size on the slit. Five spatial pixels with indices 25 to 29 were electronically windowed to record the photon counts. Each pixel has a length of 1.0 mrad along the spatial dimension (length from upper left to lower right in Figure 1a) and the low resolution slit width was set to 1.5 mrad (width from upper right to lower left in Figure 1a). During the stellar occultation, the spacecraft navigation system controls pointing through a stellar/solar referenced 3-axis stabilized platform with a deadband of 0.5 mrad referenced to the UVIS FUV principal axis (Pilinski & Lee, 2009). The thrusters of the spacecraft attitude control system react only at the deadband limit. Therefore, the star image shows motion on the slit (green line in Figure 1a), which results in changes in photon count distributions among spatial pixels along the spatial dimension and spectrum distortions along the spectral dimension. The effect of star image motion along the spatial dimension can be eliminated by summing up the photon counts received by all five spatial pixels after flat fielding. The windowed spatial pixels fully contain the stellar image during the entire T52 occultation. Motion of the star image along the spectral dimension, however, must be modeled to analyze the data, which is the focus of this work.

The star image position on the slit along the spectral dimension as a function of the ray tangent height in Titan's atmosphere during T52 is shown in Figure 1b. The star image moved at a constant rate within ±0.5 mrad between the deadband points, showing nominal function of the control system. This introduces strong distortions in the spectral dimension, as demonstrated in Figure 1c by the simulated spectra with different pointing motions. An extinction feature of $C_2H_2$ near 1520 Å is shifted by a few pixels due to the pointing motion, causing a difference in spectral structure between spectra that prevents constructing extinction spectra by dividing one spectrum by another. Moreover, due to the photon scattering effects in the instrument, a single spectral line is spread across a wide range of wavelengths, necessitating simultaneous modeling of a large swath of the spectrum. The full width at half maximum of a typical spectral line is ~1.5 Å, or approximately two spectral pixels, while the line wings extend hundreds of angstroms. This results in non-linearity in the spectrum. Figure 1d shows an example of internal photon scattering. Photon counts shortward of 1360 Å are mostly the result of internal scattering from longer wavelengths when $CH_4$ extinction saturates this range. Therefore, it is not feasible to shift the spectrum back even if the value of the pointing motion were known. A forward instrument model is essential for extracting information from these distorted spectra.

## 3 Methodology

A detailed description of Cassini/UVIS data calibration and reduction is given in Chapter 9 and 10 of the Cassini/UVIS Users Guide, available in NASA-PDS (2017). To improve statistical accuracy, spectra obtained during T52 are integrated over time intervals of 17.5 s, covering an altitude range of ~25 km. The selected integration intervals are merged with pointing information obtained from the SPICE C-kernel (NASA NAIF), which is shown in Figure 1b as the green path. The reference spectrum of the star ($I_0$) is constructed by averaging the spectra when the ray tangent height is greater than 1500 km above Titan's surface, where extinction by Titan's atmosphere is negligible, and the pointing motion is less than 0.125 mrad (half of a pixel width).

Our forward model combines an extinction model and an instrument simulator. The extinction model computes the intensity spectrum received by Cassini/UVIS, while the instrument simulator (Shemansky et al. 2005; Shemansky & Liu 2012) generates photon count observations based on the intensity spectrum, instrument internal scattering and pointing motion. The instrument simulator contains high resolution response functions for each pixel, which encapsulate the effects of instrument internal scattering that were measured in the lab and calibrated in flight. The core of the point spread function has a full width at half maximum of ~1.7 Å and the wings extend over a spectral range of 800 Å.

**Table 2**. Extinction cross sections.

| Species | Reference | Wavelengths (Å) |
|---------|-----------|-----------------|
| $CH_4$ | Kameta et al. (2002) | 1115-1426 |
| | Chen & Wu (2004) | 1426-1490 |
| $C_2H_2$ | Wu et al. (2001) | 1115-1912 |
| $C_2H_4$ | Wu et al. (2004) | 1115-1912 |
| $C_2H_6$ | Au et al. (1993) | 1115-1193 |
| | Wu et al. (2004) | 1199-1528 |
| HCN | Nuth & Glicker (1982) | 1115-1300 |
| | Lee (1980) | 1300-1568 |
| $C_4H_2$ | Ferradaz et al. (2009) | 1150-1912 |
| $C_6N_2$ | Connors et al. (1974) | 1115-1912 |
| $C_6H_6$ | Pantos et al. (1978) | 1115-1912 |
| $HC_3N$ | Ferradaz et al. (2009) | 1150-1560 |
| $C_2N_2$ | Nuth & Glicker (1982) | 1115-1701 |

Eleven hydrocarbon and nitrile species ($CH_4$, $C_2H_2$, $C_2H_4$, $C_2H_6$, HCN, $C_4H_2$, $C_6N_2$, $C_6H_6$, haze, $HC_3N$ and $C_2N_2$), which have extinction features in the FUV, are considered in the forward model. We include two species in the retrieval, $C_2N_2$ and $C_6N_2$, which have not been detected previously, to examine the extent to which their abundances can be constrained, as photochemical models suggest their existence (e.g. Willacy et al. 2016). The cross sections of these species are obtained from laboratory work (Table 2), some of which were conducted at room temperature. The differences in temperatures between that of the measurements and ambient conditions in Titan's atmosphere may contribute ~20% uncertainty to the LOS abundances. Pressure effects are negligible. Haze particles are assumed to be spherical with a radius of 12.5 nm and the same optical properties as their laboratory analogs ("tholin"; Khare et al. 1984), in line with Liang et al. (2007) and Koskinen et al. (2011). We combine the LOS abundances and cross sections to construct spectra based on a normalized $I_0$.

Retrieval is a multivariable inverse problem. Combining a proper retrieval algorithm with a forward model, physical properties can be derived from the observations. The MCMC method is used in this work to solve the inverse problem. It searches parameter space with the ability to extract asymmetric posterior probability density functions (PDFs) in a computationally feasible way. For each proposed parameter set, a spectrum is constructed with the procedure described above, and is then compared with the observed spectrum to determine the posterior probability of this parameter set. The cost function is defined as follows:

$$\ln(p) = -\frac{1}{2}\sum_i \left[ \frac{\left(I_{Obs_i} - I_{MCMC_i}\right)^2}{\sigma_i^2 + 0.1} + \ln(\sigma_i^2 + 0.1) \right]$$

where p is the posterior probability of one proposed parameter set; $I_{Obs}$ and $I_{MCMC}$ are the photon counts from the observation and those calculated from the forward model during one MCMC attempt, respectively; and $\sigma_i$ is the standard deviation of the spectral intensity at wavelength i, assumed to be the square root of the simulated photon count. A softening factor of 0.1 is added to the standard deviation at each wavelength to avoid dividing by zero when the intensity decreases to zero at some wavelengths at low altitudes. An example of $I_{Obs}$ and $I_{MCMC}$ is shown in Figure 2 as blue and green lines, respectively. With this cost function, we use the emcee package (Foreman-Mackey et al. 2013) to conduct the MCMC parameter search. An MCMC procedure with 120 chains, selected according to the number of parameters, is used to search through parameter space. A bounded uniform prior in log space is set for each parameter, so abundances are retrieved with no prior knowledge within 2 orders of magnitude of the predicted values from the latest results of the Caltech/JPL photochemical model KINETICS (Li et al. 2014, Li et al., 2015, Willacy et al. 2016). The bounds are adjusted by 2 orders of magnitude if necessary after each 2000 steps according to the PDF of the parameter and the converging criterion. The MCMC procedure is extended for 2000 more steps after the final bound adjustment to generate the resulting PDFs.

## 4 Synthetic spectrum analysis

We analyze a synthetic spectrum (red line in Figure 2) to test the reliability of our method. We use the LOS abundances of all species obtained from fitting to the observation (green line in Figure 2) as well as the pointing motion (Figure 1b) to construct the synthetic spectrum. We then add noise on the level of the square root of the simulated photon count plus the softening factor to obtain the final synthetic spectrum (red line in Figure 2) with the same calibration procedure as the data. In other words, the synthetic spectrum is the same as the simulated spectrum used to fit the observations, except it includes noise. Figure 2 shows that the noise level is slightly greater than the disagreements between the simulated spectrum and the observation, such that the analysis of the synthetic spectrum should give a lower limit to the reliability of our method.

Figure 3 shows the PDFs of the LOS abundances of hydrocarbon and nitrile species resulting from the synthetic spectrum analysis. The LOS abundance values used to construct the synthetic spectrum are shown as black dashed lines. The constraint on each LOS abundance is interpreted by fitting the PDFs with three types of functions: Gaussian, sigmoid, and constant, and the PDFs are categorized as such by comparing the residuals of the fit. LOS abundances with Gaussian-like PDFs are defined as well constrained (e.g., $CH_4$, $C_2H_2$, $C_2H_4$, HCN, $C_4H_2$, $C_6H_6$ and $HC_3N$ in Figure 3). These species typically have distinct spectral features (Figure S2) that allow for strong constraints. The retrieved LOS abundances of these species are all within ~1-σ of the true values, indicating that our retrieval method is stable to random noise. The LOS abundance PDFs of some other species (e.g., $C_2H_6$, $C_6N_2$, haze and $C_2N_2$ in Figure 3) show asymmetric behavior, and so can only provide upper limits through fits to the sigmoid function. Interestingly, ethane ($C_2H_6$), one of the major hydrocarbons, belongs in this category due to overlapping spectral features with the most abundant hydrocarbon, methane ($CH_4$, Figure 1d), whose LOS abundance is two orders of magnitude higher, resulting in the anti-correlation of these two PDFs (Figure S3). The failure to retrieve ethane is consistent with previous results obtained above 700 km during flyby T41i (Koskinen et al. 2011). In some cases, the PDFs may be between Gaussian and sigmoid, such as that of $C_6N_2$ in this analysis, which shows a peak with poorly constrained lower limits; as the type of PDF is still determined by the value of the residuals, the case of $C_6N_2$ and others like it would either have only upper limits (more sigmoid-like) or be constrained with relatively large uncertainties (more Gaussian-like).

## 5 Results and discussion

An example spectrum from the T52 occultation observation at ~750 km is shown in Figure 2, together with a simulated spectrum that best fits the observation. PDFs for the LOS abundances of hydrocarbon and nitrile species resulting from retrievals of this spectrum are shown in Figure 4. Most of the behavior of these PDFs are consistent with those in Figure 3. Six species ($CH_4$, $C_2H_2$, $C_2H_4$, $C_4H_2$, $C_6H_6$ and $HC_3N$) show Gaussian-like PDFs, and are thus well constrained with precise values for the LOS abundances and small uncertainties. It is worth noting that the uncertainties shown here are only from photon noise; another probable source of uncertainties is the presently unavailable temperature dependencies of extinction cross sections, which may introduce systematic errors. Three other species ($C_2H_6$, $C_6N_2$, and $C_2N_2$) have asymmetric PDFs more similar to the sigmoid function, with only well-defined upper limits. HCN and haze both have PDFs that are close to Gaussian, and as such are categorized as well constrained, but with large uncertainties (about a factor of 3). The third PDF type, a constant, is not shown in either

Figure 3 or Figure 4, as all retrieved species are constrained to some extent at ~750 km. A constant PDF usually takes place only when the ray tangent height is either too high or too low in the atmosphere, where the LOS abundances of some species are either insufficient to be seen or overwhelmed by saturated absorption, respectively. In other words, flat PDFs would be returned when there is almost no information in the spectrum.

Repeating the procedures outlined above for all available altitudes, we present the vertical profiles of LOS abundances of all eleven species in Figure 5 with error bars and arrows for well-defined constraints and upper limits, respectively. This is the first time that the LOS abundance profiles of these species are retrieved from a flyby with significant pointing motion. Comparison of the $CH_4$, $C_2H_2$, $C_2H_4$ and HCN profiles with those retrieved from the TB (Shemansky et al. 2005) and T41i flybys (Koskenin et al. 2011) are given in Figure S5. It shows general agreement between these three retrievals despite some differences at higher altitudes, which may result from the different seasons and/or latitudes. Among the major hydrocarbons (Figure 5a), only the most abundant species, $CH_4$, is detectable above 1000 km, while all the others only exhibit upper limits. With decreasing altitude, concentrations of larger organic molecules increase due to photochemistry. All of the major hydrocarbons except $C_2H_6$ remain well constrained down to 400 km, where absorption saturates. Spectral features of $C_2H_6$ overlap with that of $CH_4$ for most of the altitude range, resulting in a failure to retrieve its abundance, as discussed in Section 4. Among nitriles, HCN is constrained with relatively large uncertainties between 700 and 1000 km, while $HC_3N$ is well constrained to much lower altitudes. Only upper limits for $C_6N_2$ and $C_2N_2$ are obtained, and so we do not claim a detection. Tighter constraints may be obtained from spectra with higher signal to noise ratio measured during a more stable flyby (e.g., T41i), which we will investigate in a future publication. Aside from the major hydrocarbon and nitriles, our new method can also identify the two long wavelength absorbers, benzene and haze. Benzene is well constrained below 900 km due to its distinguishable feature near 1790 Å, while haze particles can only be retrieved below 750 km. The range of altitudes where haze is well constrained in our work is smaller than those of Liang et al. (2007) and Koskinen et al. (2011). Liang et al. (2007) retrieved the haze profile up to 1000 km from observations obtained during flyby TB by assuming that all extinction between 1850 to 1900 Å is caused by haze, as benzene was not included as a potential absorber. However, the current work shows that benzene can contribute up to 50% of the extinction in this wavelength range, so the ambiguity of the two absorbers needs to be considered. Koskinen et al. (2011) also retrieved the haze profile up to 1000 km using observations obtained during two stable flybys T41i and T53, while assuming that the PDF of the LOS abundance of haze was Gaussian. Both the higher signal to noise ratio of these observations and the assumption of a Gaussian PDF could have contributed to a greater altitude range where haze LOS abundance can be constrained. Applying our method to these flybys may help to understand why these differences exist.

LOS abundances of eight species ($CH_4$, $C_2H_2$, $C_2H_4$, HCN, $C_4H_2$, $C_6H_6$, haze, and $HC_3N$) are converted to number density profiles to allow for ease of comparison to photochemical models. Three species ($C_2H_6$, $C_2N_2$ and $C_6N_2$) are excluded, as their LOS abundances are not well constrained over most of the considered altitude range. The Abel inverse transform is used here, which assumes spherical symmetry, to compute the vertical profiles. A Bootstrap Monte Carlo (BSMC) method is used to evaluate the quality of conversion and provide uncertainties. BSMC has the advantage of being applicable to different types of PDFs of the LOS abundances, which is necessary since a number of the PDFs are not Gaussian. The number density profiles of each species are computed individually since each species is independent of others. In each computation, a set of LOS abundances at all retrieving altitudes for the given species is sampled from their PDFs (e.g. one of the panels in Figure 4) at each BSMC step, from which we compute the corresponding vertical number density profile. As the distribution of species LOS abundances in the Markov chains obtained from the retrieval are identical to their PDFs when the MCMC procedure reaches equilibrium, we used the values in the last 1500 steps of each of the 120 chains as the sampling procedure for BSMC and generate 180000 number density profiles for each species. Therefore, at each altitude of an individual species, we obtain 180000 probable number densities, which form a number density PDF. To interpret these PDFs, we use the same method of fitting them with three types of functions (Gaussian, sigmoid, and constant) and categorized them by comparing the residuals as mentioned in Section 4. Number density profiles corresponding to well-constrained number densities and upper limits with positive values are shown in Figure 6. As the number density at each altitude is computed with contributions from both well and poorly constrained LOS abundances, the number densities have larger relative uncertainties and smaller ranges of altitudes where they are well constrained. Major hydrocarbons with large abundances and distinct spectral features ($CH_4$, $C_2H_2$ and $C_2H_4$; Figure 6a-6c) are well constrained over a wide range of altitudes. In contrast, the number density of the most abundant nitrile, HCN, is poorly constrained for most of the altitude range considered (Figure 6d) due to its large LOS abundance uncertainty (Figure 5). Other minor species (Figure 6e-6h) are well constrained over at least some of the considered altitude range, and can thus provide constraints on Titan's atmospheric chemistry.

6 Conclusions

A new method to correct for the effects of pointing motion of Cassini/UVIS has been developed using an instrument simulator and the MCMC method. The new approach is successfully applied to the T52 stellar occultation observations of Titan's atmosphere to retrieve the LOS abundances and number densities of hydrocarbon and nitrile species, and allows for the quantification of how well each LOS abundance and number density can be constrained, facilitating the analysis of all Cassini/UVIS stellar occultations at Titan. Application of the present method to all available observations is expected to reveal seasonal and latitudinal variations in the atmospheric composition of Titan, thereby providing useful constraints for photochemical and global circulation models.

**Acknowledgments**

This research was supported in part by the Cassini/UVIS program via NASA grant JPL.1459109 to the California Institute of Technology, and was partially supported by funding from NASA's Astrobiology Institute's proposal "Habitability of Hydrocarbon Worlds: Titan and Beyond" (PI R.M. Lopes). All the data and tools in this work are publicly available. Cassini/UVIS data is available on NASA PDS (pds.nasa.gov). The python package emcee is available at dfm.io/emcee/current. We thank Dr. Tommi T. Koskinen and Dr. Karen Willacy for sharing results.

We attach the following files as supporting information:

Figure S1. An extended version of Figure 1c showing the spectra in the entire FUV wavelength range.

Figure S2. The spectral contribution of each species to the simulated spectrum shown in Figure 2.

Figure S3. An extended version of Figure 3 showing the correlations among parameters.

Figure S4. An extended version of Figure 4 showing the correlations among parameters.

Figure S5. Comparison of LOS abundances of $CH_4$, $C_2H_2$, $C_2H_4$ and HCN with two previous works (TB and T41i).

T52_LOS_abund.csv. A list of comma-separated values (CSV) of the LOS abundances derived in this work and shown in Figure 5.

T52_Num_Dens.csv. A list of comma-separated values (CSV) of the number density profiles derived in this work and shown in Figure 6.

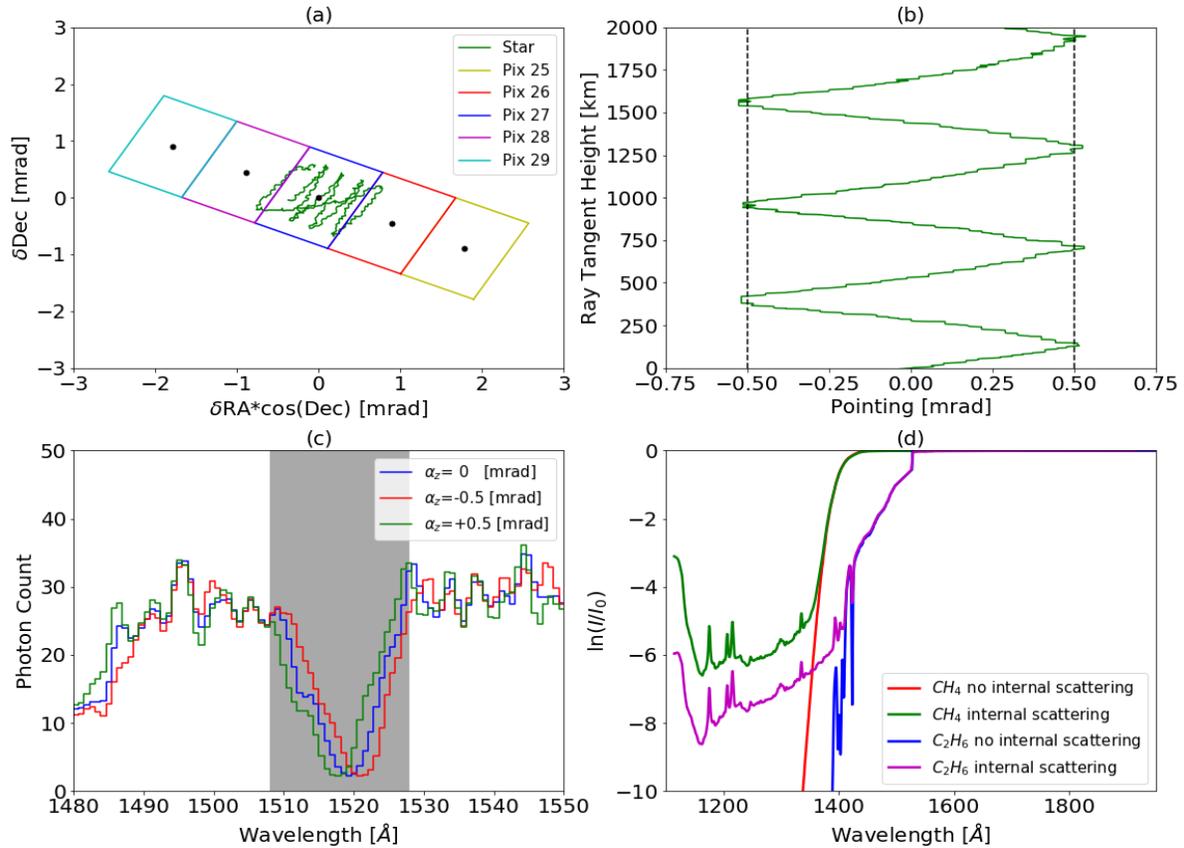

**Figure 1**. Pointing motion and its effects. (a) Spatial orientation of Cassini/UVIS detector pixels (color rectangles) with their centers (black points) and the star motion on the detector (green solid line) obtained from the SPICE C-kernel. The pixel with index 27 (blue rectangle) is the center pixel during the T52 occultation observation. The spatial dimension is along the length of the 5 pixels, with the spectral dimension being perpendicular to that. (b) Vertical profile of pointing motion above Titan's surface along the spectral dimension (solid green line). Deadbands along this dimension at each end are denoted by two vertical black dashed lines. (c) Simulated photon count spectra from the T52 occultation at 754 km ray tangent height with pointing values along the spectral dimension of 0 (blue), -0.5 (red) and +0.5 mrad (green). The shaded area indicates the signature extinction feature of $C_2H_2$ near 1520 Å. Figure S1 shows the full extent of the three spectra in the FUV wavelength range. (d) Extinction spectra of $CH_4$ and $C_2H_6$ with (green for $CH_4$ and magenta for $C_2H_6$) and without (blue for $CH_4$ and red for $C_2H_6$) taking into account instrument internal scattering. The LOS abundances for both species are set to $10^{18}$ cm$^{-2}$, which is approximately the value for that of $CH_4$ at 750 km.

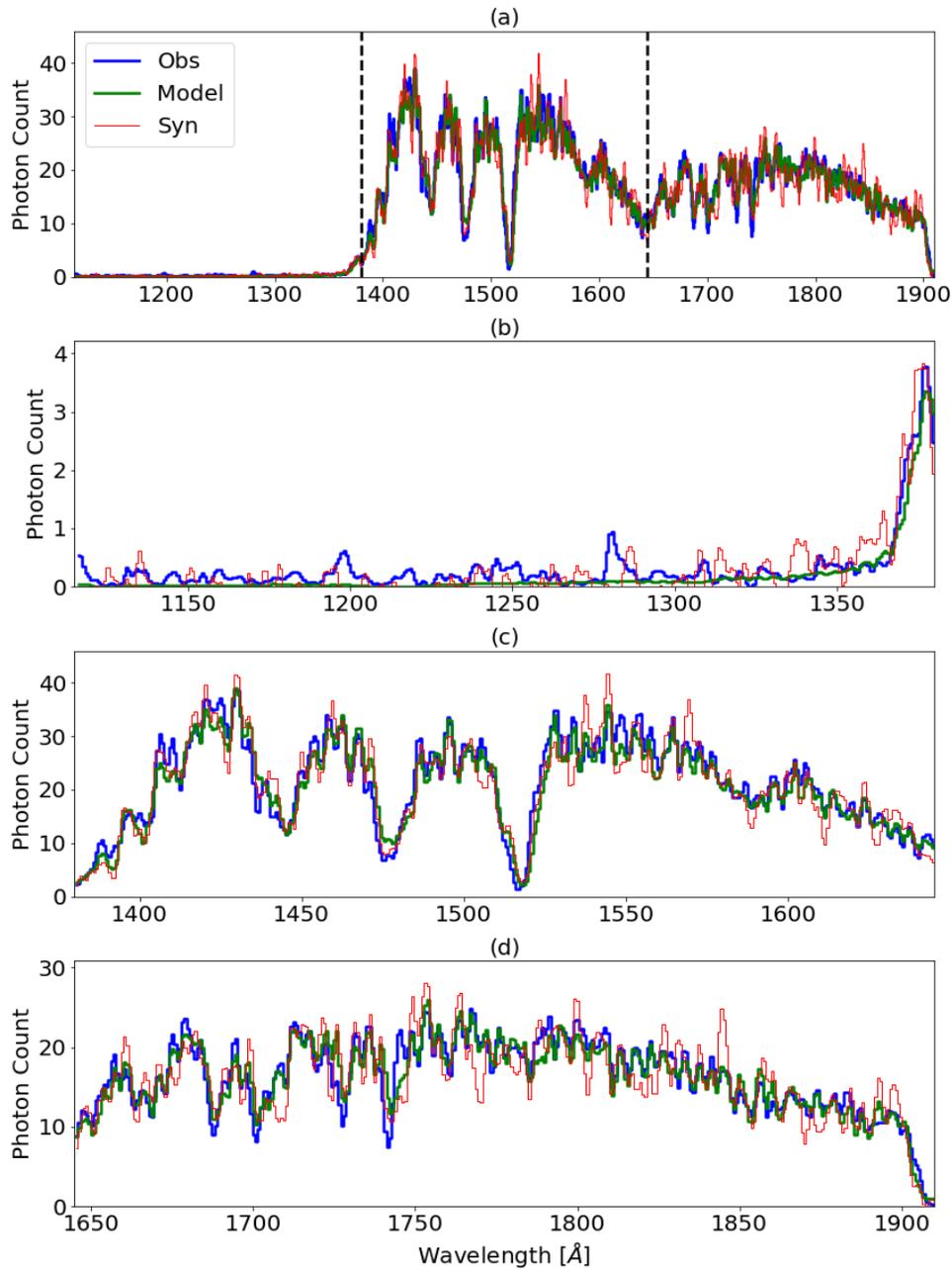

**Figure 2**. **(a)** Photon count spectra of the T52 occultation at 754 km ray tangent height showing the observed spectrum (blue), the best fit simulated model spectrum (green) and the synthetic spectrum (red), which includes artificially introduced noise. (b)-(d) Detailed views of the photon count spectra split along the black dashed lines in (a). The y axes scales are different in (b)-(d) for the purpose of presentation. The spectral contribution of each species to this spectrum is shown in Figure S2 in the supporting information.

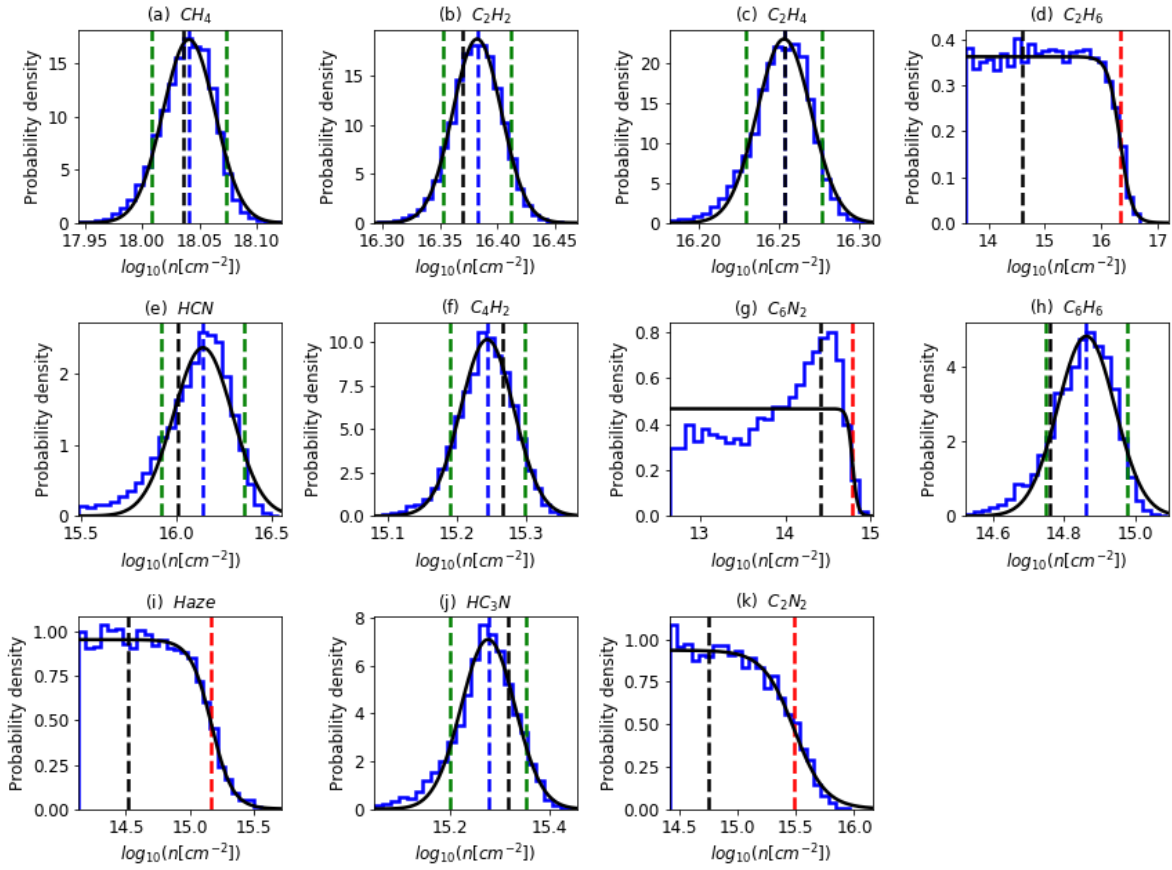

**Figure 3**. Probability density functions (PDFs, blue solid lines) of the logarithm of LOS abundances retrieved from the synthetic spectrum (red line in Figure 2). The black dashed lines indicate the LOS abundances used to generate the synthetic spectrum, i.e. the "true" value. The best fit function to each PDF (Gaussian, sigmoid, and constant) is shown as black solid lines. The median and 1-σ confidence interval are denoted by blue and green dashed lines, respectively, for well-constrained species. For others, the upper limits are denoted by red dashed lines. The correlations among the parameters are shown in Figure S3 in the supporting information.

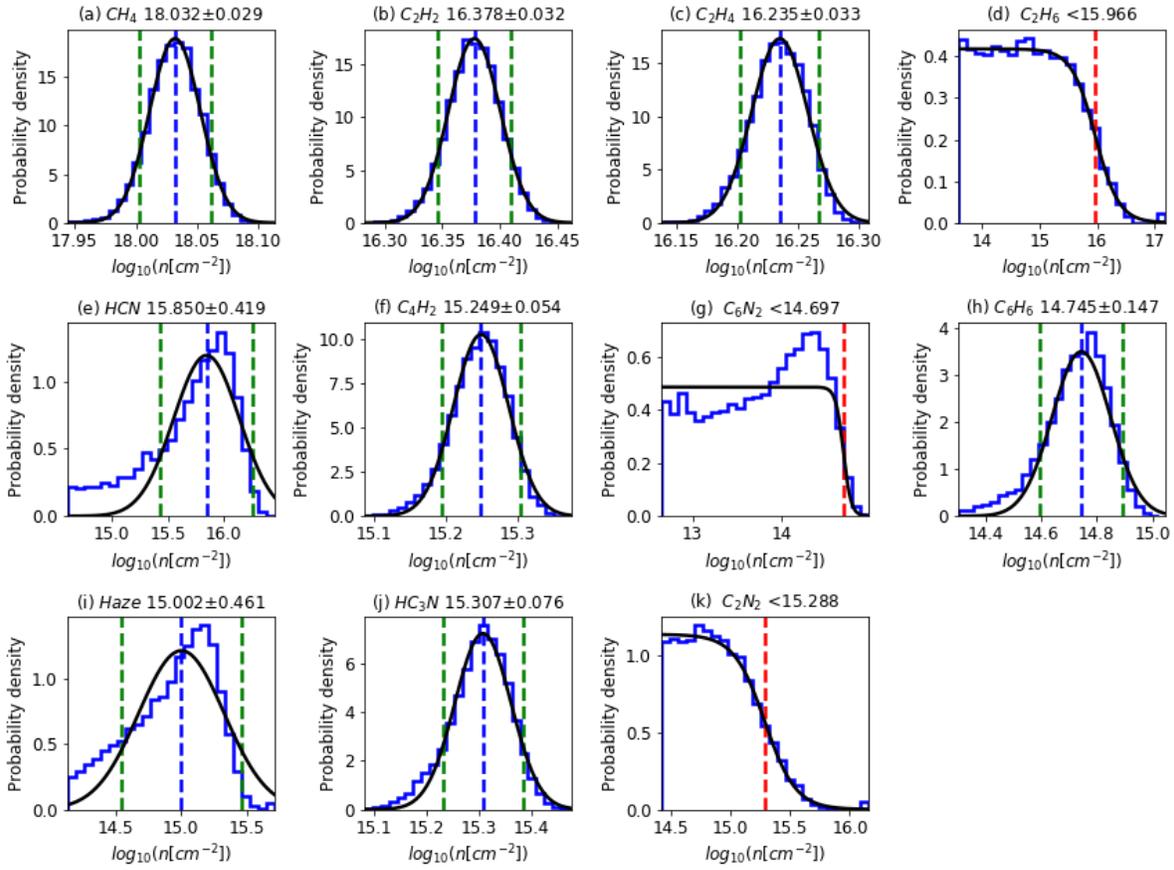

**Figure 4**. Probability density functions (PDFs) of the logarithm of LOS abundances retrieved from a photon count spectrum at 754 km ray tangent height (blue solid lines). The best fit function to each PDF (Gaussian, sigmoid, and constant) is shown as black solid lines. The median and 1-σ confidence interval are denoted by blue and green dashed lines, respectively, for well-constrained species. For others, the upper limits are denoted by red dashed lines. The medians, 1-σ confidence intervals, and upper limits are also given atop each subplot. The correlations among the parameters are shown in Figure S4 in the supporting information.

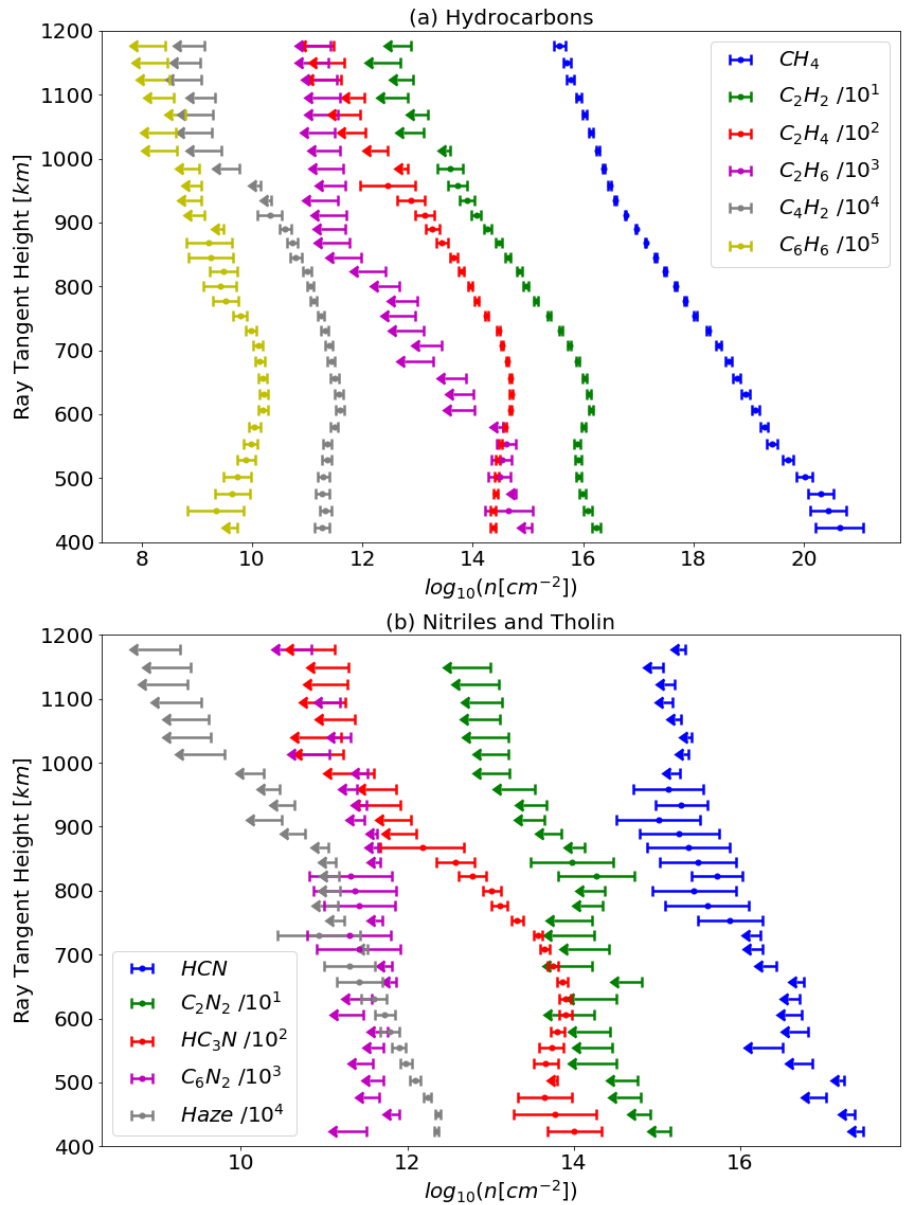

**Figure 5**. Vertical profiles of the logarithm of LOS abundances retrieved from T52 occultation observations. Some species are offset by a few orders of magnitude for the purpose of presentation. Points with error bars denote well-constrained values, while arrows denote upper limits. The lengths of the arrows denote the width of each soft upper limit threshold. Haze particles are assumed to be 12.5 nm spheres with the same optical properties as their laboratory analog ("tholin"; Khare et al. 1984). Data used to generate this figure is available in the supporting information. Comparison of LOS abundances of $CH_4$, $C_2H_2$, $C_2H_4$ and HCN with two previous flybys (TB and T41i) is given in Figure S5.

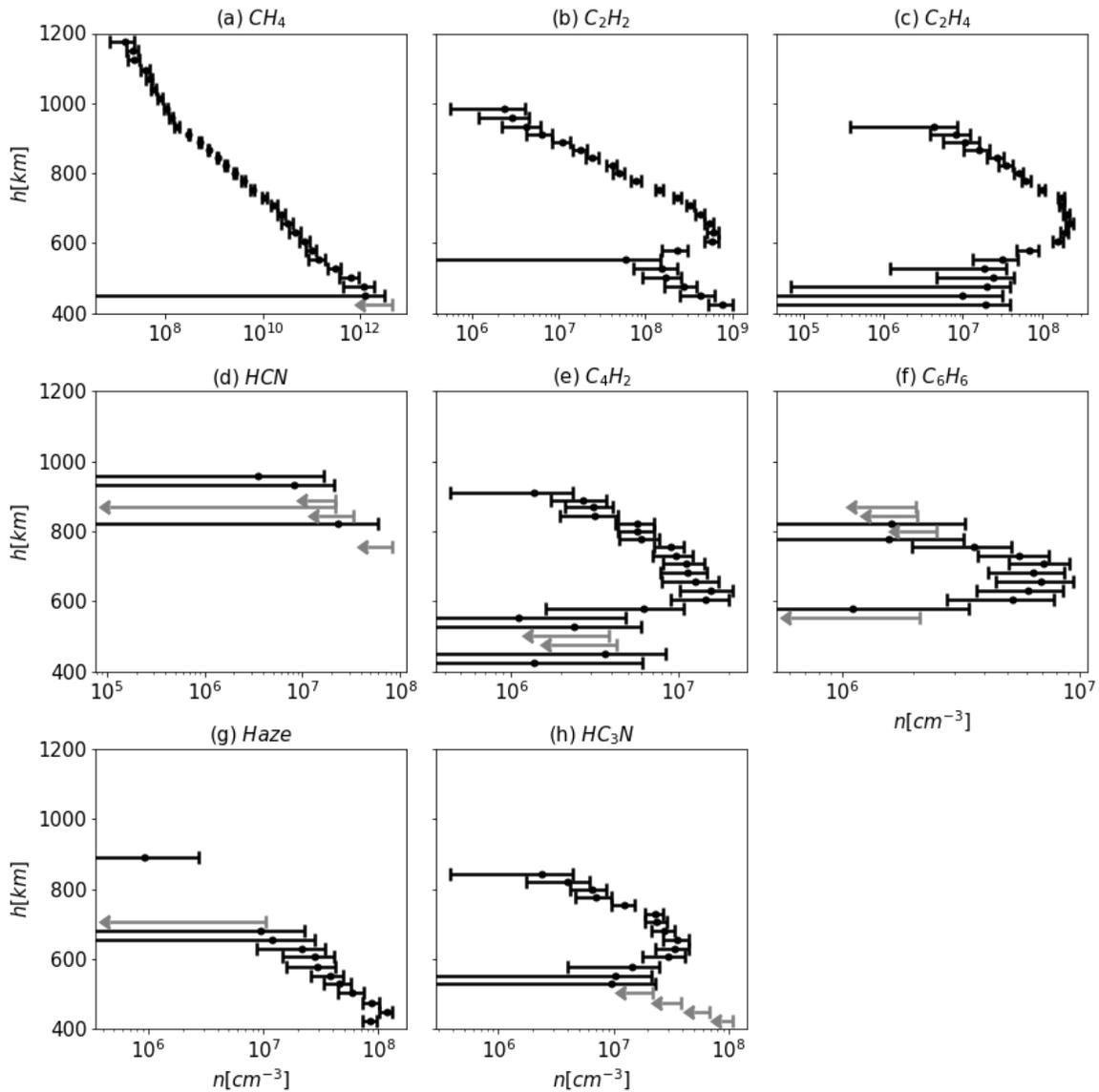

**Figure 6.** Number density profiles of selected species retrieved from T52 occultation observations. Black points with error bars denote well-constrained values, while grey arrows denote upper limits. The lengths of the arrows denote the width of each upper limit soft threshold. Haze particles are assumed to be 12.5 nm spheres with the same optical properties as their laboratory analog ("tholin"; Khare et al. 1984). Data used to generate this figure is available in the supporting information.

**Retrieval of Chemical Abundances in Titan's Upper Atmosphere from Cassini UVIS Observations with Pointing Motion**

Siteng Fan[1], Donald E. Shemansky[2], Cheng Li[1], Peter Gao[3], Linfeng Wan[4], Yuk L. Yung[1,5]

[1]California Institute of Technology, Pasadena, CA, 91125.

[2]Space Environment Technologies, Altadena, CA, 91101.

[3]University of California, Berkeley, CA, 94720.

[4]University of California, Santa Cruz, CA, 95064.

[5]Jet Propulsion Laboratory, California Institute of Technology, Pasadena, CA, USA

**Contents of this file**

**Introduction**

The influence of pointing motion on the distortion of the photon count spectrum of T52 at 754 km ray tangent height is shown in this supporting information as Figure S1. It is the same as Figure 1c in the main text, but expanded to include the entire FUV wavelength range.

The spectral extinction contribution of each species to the spectrum of T52 at 754 km ray tangent height is shown in this supporting information as Figure S2. Even though the zero-extinction spectrum does not exhibit a pointing drift, we add that of the spectrum at 754 km to allow for better inter-comparison.

Correlations among the probability density functions (PDFs) shown in Figures 3 and 4 in the main text are plotted in this supporting information as Figures S3 and S4, respectively. Numbers and units in Figures S3 and S4 are the same as those shown in Figures 3 and 4.

The LOS abundances of $CH_4$, $C_2H_2$, $C_2H_4$, and HCN retrieved from T52 occultation observations and shown in Figure 5 are compared with those from two previous flybys (TB; Shemansky et al. 2005 and T41i; Koskinen et al. 2011) in Figure S5.

Data shown in Figures 5 and 6 are attached in this supporting information as CSV files named T52_LOS_Abund.csv and T52_Num_Den.csv, respectively. The files contain line-of-sight abundances and number density profiles of hydrocarbon and nitrile species in Titan's upper atmosphere derived in this work using Cassini/UVIS stellar occultation observations during flyby T52.

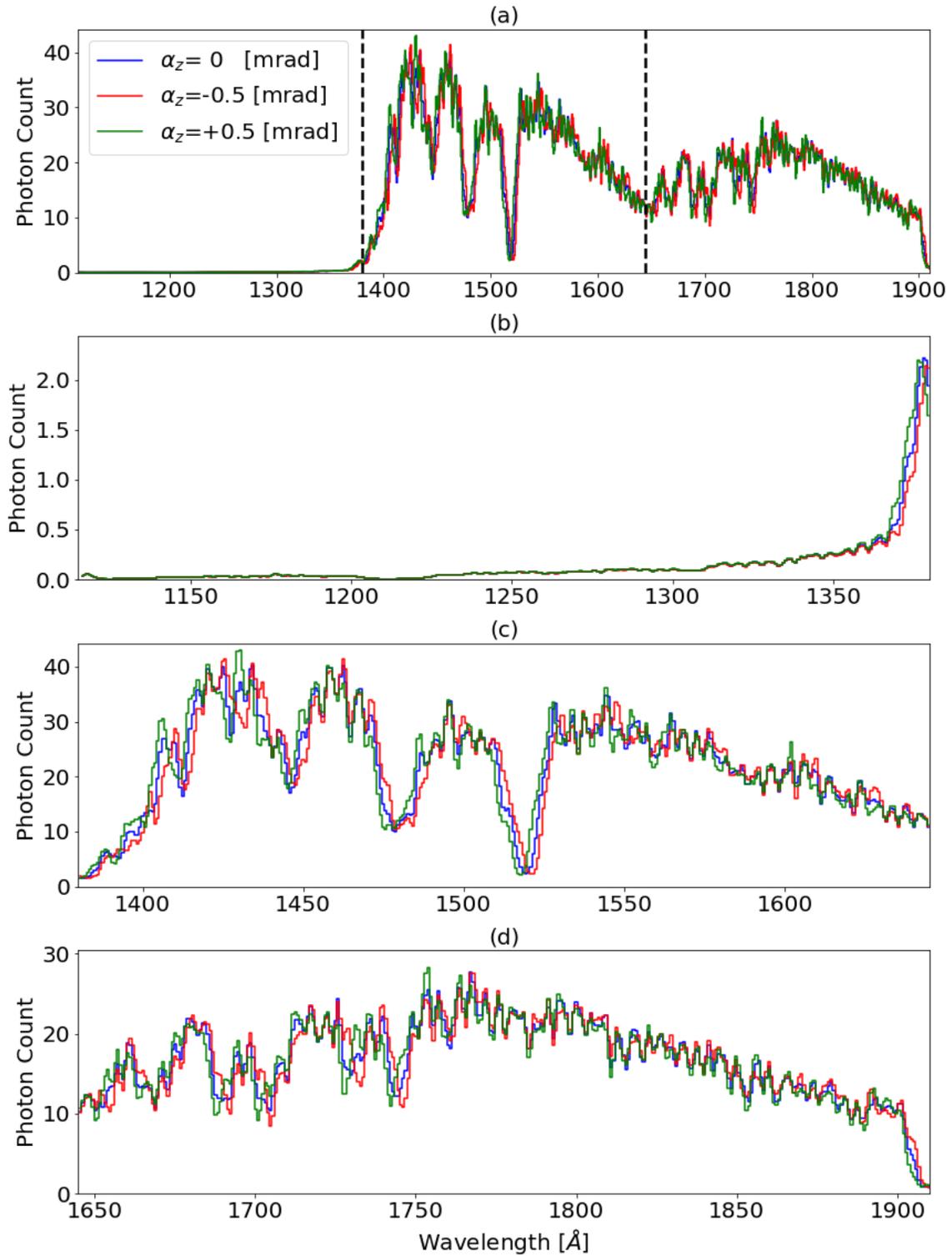

**Figure S1.** (a) Simulated photon count spectra from the T52 occultation at 754 km ray tangent height with pointing motion values along the spectral dimension of 0 (blue), -0.5 (red) and +0.5 (green) mrad. (b)-(d) Detailed views of the photon count spectra split along the black dashed lines in (a). The y-axes scales are different in (b)-(d).

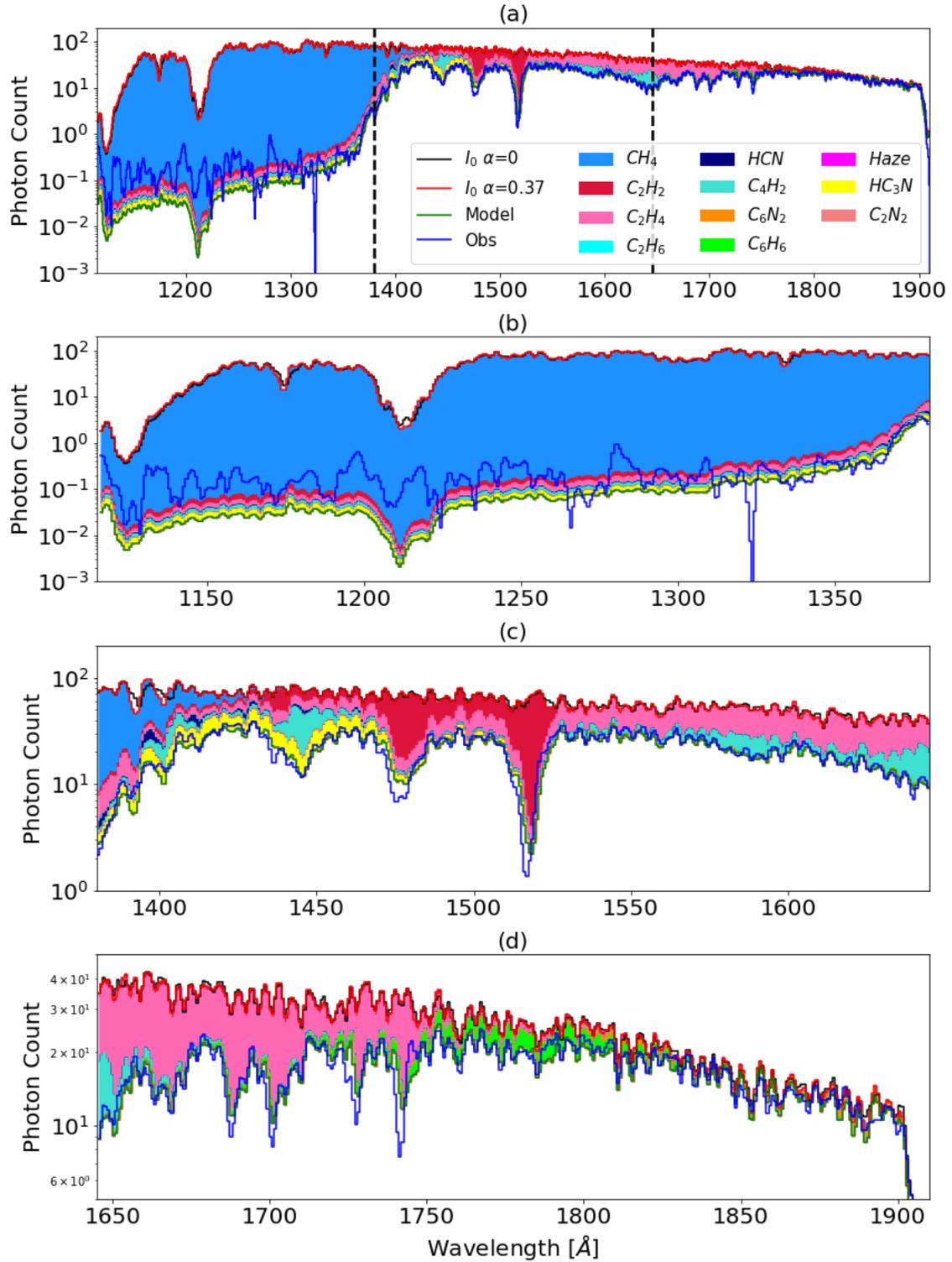

**Figure S2.** (a) Photon count spectra from the T52 occultation. The spectra observed above Titan's atmosphere ($I_0$) without and with a pointing drift of 0.37 mrad, its value at 754 km ray tangent height, are shown as black and red solid lines, respectively. The observed and best-fit model spectra at 754 km ray tangent height are shown as blue and green solid lines,

respectively. The extinction contributions of each species are shown by the shaded areas between spectra. The y-axis is in log scale such that the area denoted for each species is proportional to their contributions to optical depth. (b)-(d) Detailed views of the photon count spectra split along the black dashed lines in (a). The y-axes scales are different in (b)-(d).

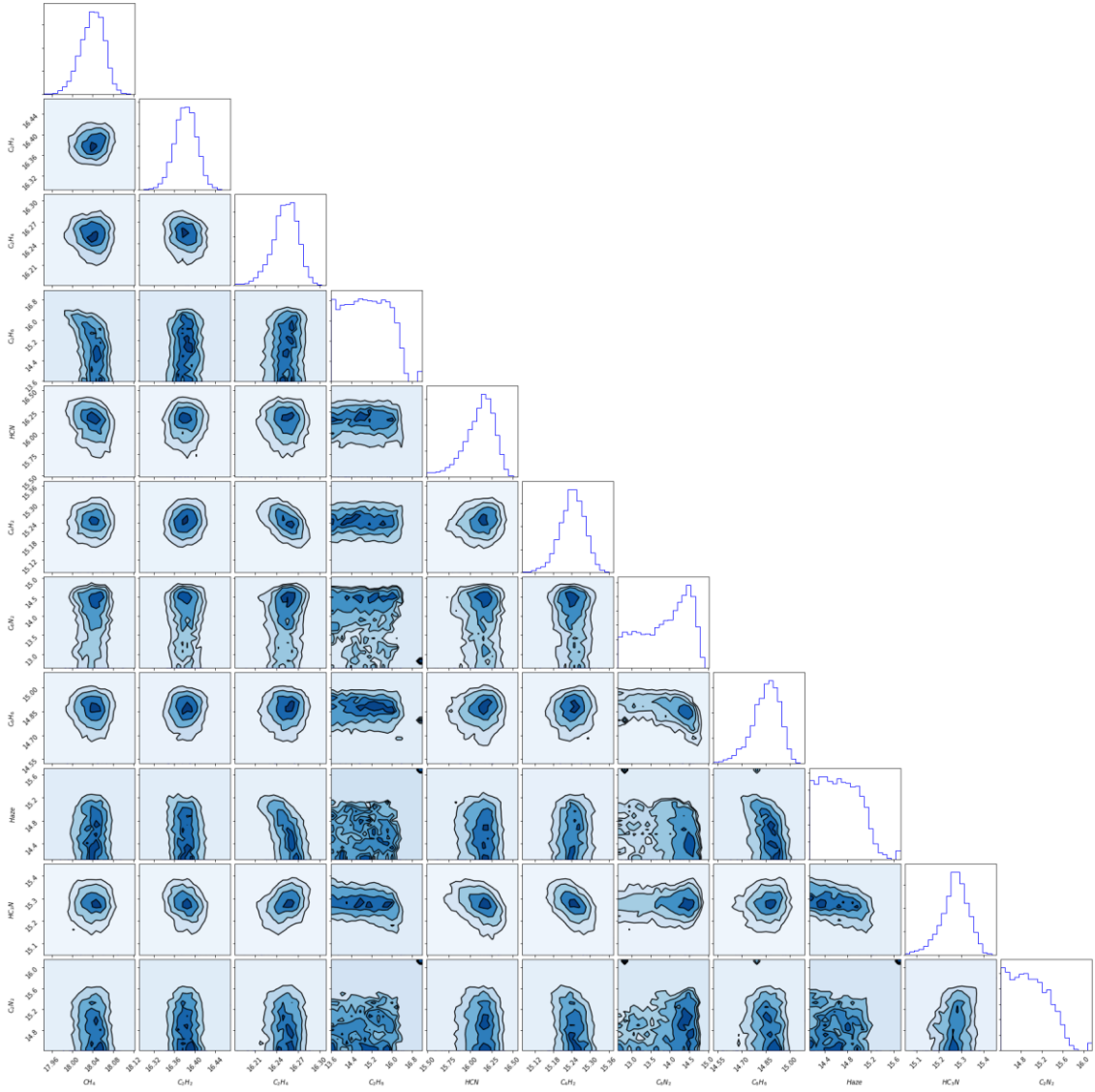

**Figure S3.** Probability density functions (PDFs) of the logarithm of LOS abundances, which are the same as those shown in Figure 3, and the correlations among them retrieved from the synthetic spectrum shown in Figure 2. Axes units are the same as those in Figure 3.

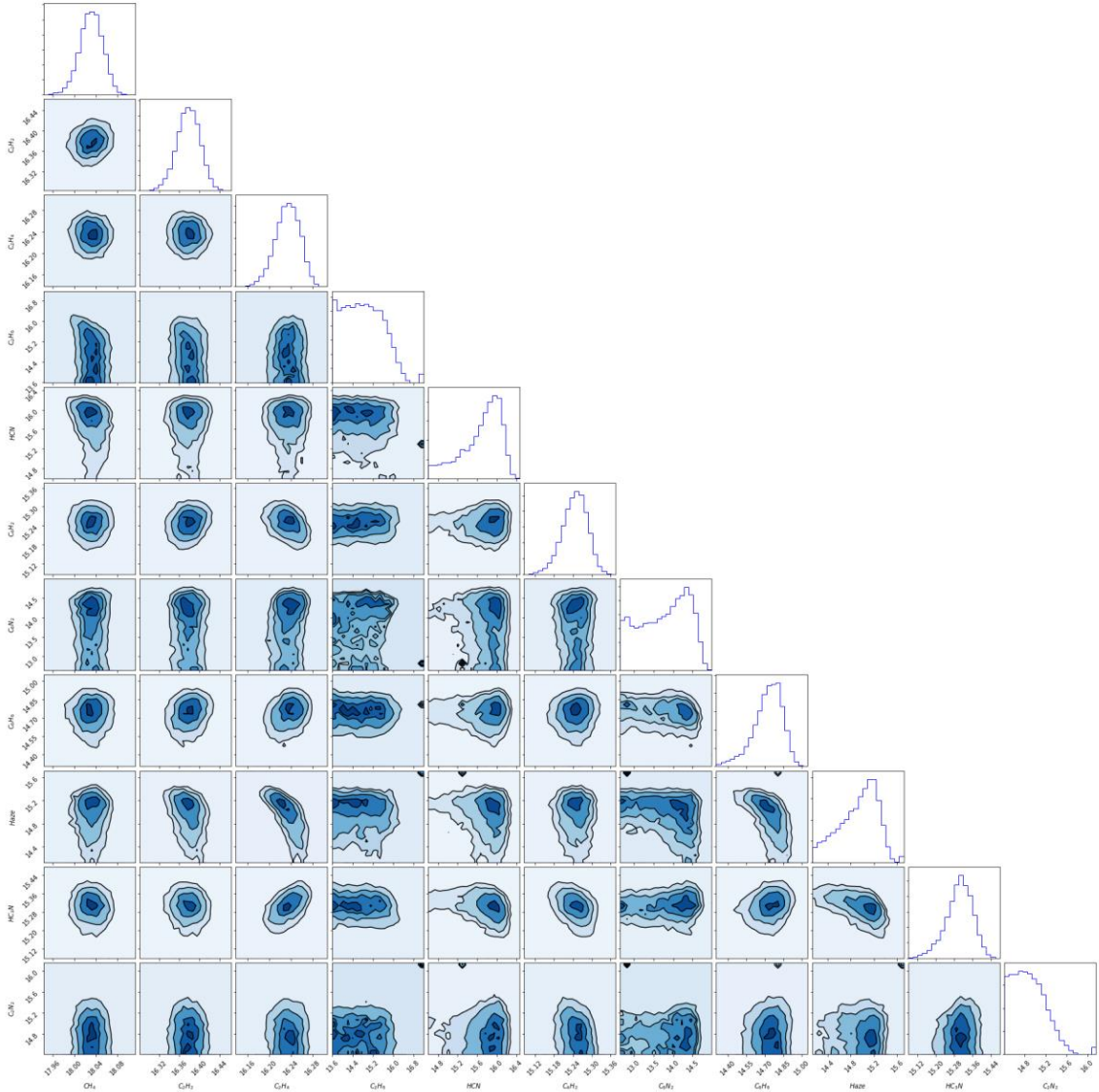

**Figure S4**. Probability density functions (PDFs) of the logarithm of LOS abundances, which are the same as those shown in Figure 4, and the correlations among them retrieved from the observed photon count spectrum at 754 km ray tangent height shown in Figure 2. Axes units are the same as those in Figure 4.

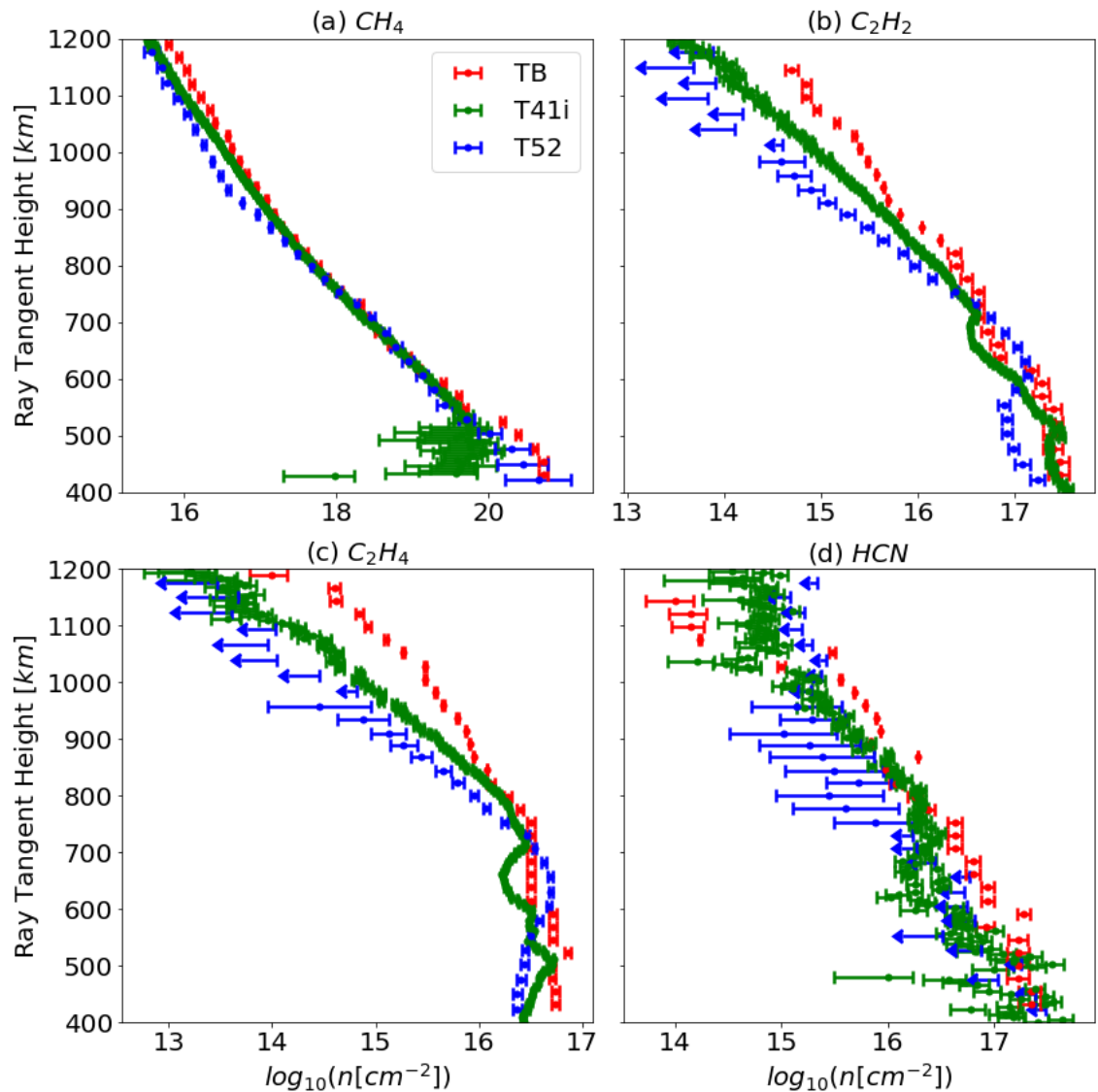

**Figure S5.** Vertical profiles of the logarithm of LOS abundances retrieved from T52 occultation observations (blue), compared to those retrieved from observations during TB (red; Shemansky et al. 2015) and T41i (green; Koskinen et al. 2011) by the cited works.

**Dataset S1.** Line-of-sight abundances of hydrocarbon and nitrile species in Titan's upper atmosphere derived from our retrievals. The dataset contains 34 columns of comma-separated values. The first column is altitude in km, which is the ray tangent height above Titan's surface. Columns 2 to 34 are denoted [Species name]_[Attribute name], where [Attribute name] can take on values of "Flag", "Value", and "Width". The meanings of these attributes are as follows:

1. [Species name]_Flag: quality flag for the LOS abundance of this species.
   0: well constrained with a value and error (shown as points with error bars in Figure 5), where the PDF can be fit by $f(x)=a*exp(-(x-V)^2/W^2)$, where x is the x-axis of the PDF, a is an amplitude, V is given in the corresponding "Value" column, and W is given in the corresponding "Width" column.
   1: loosely constrained with an upper limit and width (shown as arrows in Figure 5), where the PDF can be fit by $f(x)=a*(1+tanh(-(x-V)/W))$.
   2: no constraints.
2. [Species name]_Value: logarithm of the LOS abundance of this species in $cm^{-2}$.
   When Flag=0: value of LOS abundance.
   When Flag=1: value of upper limit.
   When Flag=2: no meaning.
3. [Species name]_Width: absolute uncertainty of value V in logarithm of the LOS abundance of this species in $cm^{-2}$.
   When Flag=0: uncertainty of the value V of LOS abundance in log space.
   When Flag=1: width of the sigmoid function (width of the soft threshold) in log space.
   When Flag=2: no meaning.

**Dataset S2.** Number density profiles of hydrocarbon and nitrile species in Titan's upper atmosphere derived from our retrievals. The dataset contains 25 columns of comma-separated values. The first column is altitude in km, which is the ray tangent height above Titan's surface. Columns 2 to 25 are denoted [Species name]_[Attribute name], where [Attribute name] can take on values of "Flag", "Value", and "Width". The meanings of these attributes are as follows:

1. [Species name]_Flag: quality flag for the number density of this species.
   0: well constrained with a value and error (shown as points with error bars in Figure 6), where the PDF can be fit by $f(x)=a*\exp(-(x-V)^2/W^2)$, where x is the x-axis of the PDF, a is an amplitude, V is given in the corresponding "Value" column, and W is given in the corresponding "Width" column.
   1: loosely constrained with an upper limit and width (shown as arrows in Figure 6), where the PDF can be fit by $f(x)=a*(1+\tanh(-(x-V)/W))$.
   2: no constraints.

2. [Species name]_Value: number density of this species in $cm^{-3}$.
   When Flag=0: value of number density.
   When Flag=1: value of upper limit.
   When Flag=2: no meaning.

3. [Species name]_Width: absolute uncertainty of value V in $cm^{-3}$.
   When Flag=0: uncertainty of the value V.
   When Flag=1: width of the sigmoid function (width of the soft threshold).
   When Flag=2: no meaning.